\documentclass[useAMS,usenatbib]{mn2e}
\usepackage{epsfig}

\usepackage{amssymb}

\title[Time-resolved spectroscopy of the pulsating CV GW Lib]{Time-resolved spectroscopy of the pulsating CV GW Lib}
\author[L. van Spaandonk et al.]{L. van Spaandonk$^1$\thanks{E-mail:L.van-Spaandonk@warwick.ac.uk}, D. Steeghs$^{1,2}$, T.R. Marsh$^1$, M.A.P. Torres$^2$\\
$^1$Astronomy and Astrophysics, Dept. of Physics, University of Warwick, Coventry CV4 7AL, UK\\
$^2$Harvard-Smithsonian  for Astrophysics, 60 Garden Street, Cambridge, MA 02138, USA}
\begin{document}

\date{Accepted 2009 September 21. Received 2009 July 17 }

\pagerange{\pageref{firstpage}--\pageref{lastpage}} \pubyear{2009}

\maketitle

\label{firstpage}

\begin{abstract}

We present time-resolved optical spectroscopy of the dwarf nova GW Librae during its rare April 2007 super-outburst and compare these with quiescent epochs. 
The data provide the first opportunity to track the evolution of the principal spectral features. In the early stages of the outburst, the optically thick disc dominates the optical and the line components show clear orbital radial velocity excursions. In the course of several weeks, optically thin regions become more prominent as strong emission lines replace the broad disc absorption.

Post-outburst spectroscopy covering the I-band illustrates the advantages of Ca~\textsc{ii} relative to the commonly used Balmer lines when attempting to constrain binary parameters. Due to the lower ionisation energy combined with smaller thermal and shear broadening of these lines, a sharp emission component is seen to be moving in between the accretion disc peaks in the Ca~\textsc{ii} line. No such component is visible in the Balmer lines. We interpret this as an emission component originating on the hitherto unseen mass donor star. This emission component has a mean velocity of $\sim -15\pm 5$~km~s$^{-1}$ which is associated with the systemic velocity $\gamma$, and a velocity semi-amplitude of $K_{\mathrm{em}} =  82.2 \pm 4.9$~km~s$^{-1}$. Doppler tomography reveals an asymmetric accretion disc, with the S-wave mapping to a sharp spot in the tomogram with a velocity consistent to what is obtained with line profile fitting. A centre of symmetry analysis of the disc component suggests a very small value for the WD orbital velocity $K_1$ as is also inferred from double Gaussian fits to the spectral lines. 

While our conservative dynamical limits place a hard upper limit on the binary mass ratio of $q<0.23$, we favour a significantly lower value near $q\sim0.06$. Pulsation modeling suggests a WD mass $\sim 1 M_\odot$. This, paired with a low mass donor, near the empirical sequence of an evolved CV close to the period bounce, appears to be consistent with all the observational constraints to date.

\end{abstract}


\section{Introduction}
Cataclysmic variables (CVs) are semi-detached interacting binary systems containing a white dwarf (WD) primary and a late-type main-sequence secondary (MS) which transfers mass onto the primary through Roche Lobe overflow, see \citet{warner1995} for a review. Strong and broad emission lines are a common observational feature in the majority of CVs, probing the dissipative processes within the accretion flow. The hydrogen Balmer series usually dominate the spectrum, and in the vast majority of spectroscopic studies of CVs serve as the main proxy for determining the system parameters. However, relatively unexplored spectral features remain, such as the Ca~\textsc{ii} triplet lines in the I-band. Common emission line sources include the accretion disc, the interaction point of the gas stream with the disc (hot spot)  and the irradiated face of the secondary star.

GW Lib is a dwarf nova (DN) which was discovered by \citet{gonzalez1983} as it went onto its first recorded outburst and rapidly increased in brightness by 8 magnitudes. It remained in its quiescent state after that period of activity until 2007. \citet{vanzyl2000} found three periodicities in the quiescent photometric light curve at  236, 376 and 650 s, making the primary in GW Lib the first multi-period non-radial pulsating WD in an accreting binary. \citet{thorstensen2002} measured the radial velocity curve using H$\alpha$ and determined a rather short orbital period of $76.79 \pm 0.02$~min. 
In the latest Ritter \& Kolb catalogue, only 8 systems with a smaller period are known when excluding the AM CVn systems which have H-deficient donors \citep{rkcat}, implying an evolved binary close to the period minimum. From model fits to the WD absorption troughs and mean spectra, various temperatures have been found. UV spectra suggest $T_{\mathrm{WD}} = 14,700$~K with a WD mass of $0.6 M_\odot $ \citep{szkody2002} or a dual temperature WD with $T_\mathrm{low} = 13,300$~K and $T_\mathrm{high} = 17,000$~K. The latter model can only explain the observed UV/optical pulse amplitude ratio in GW Lib (see table 2 in \citealt{szkody2002}) as compared with single pulsating ZZ-Ceti stars, if the WD mass of GW~Lib is larger than $0.8 M_\odot$. \citet{thorstensen2002} fit the optical Balmer line profiles to find $T = 13,220$~K and $\log g = 7.4$. However, all temperature estimates place the WD outside the ZZ Ceti instability strip for single WD pulsators.  A mass determination based on the detected pulsation periods combined with distance and UV flux constraints suggests a WD mass of $~1.02 M_\odot$ \citep{townsley2004}.

On 2007 April 12, amateur astronomers  reported the sudden and rapid brightening of GW Lib \citep{templeton2007} indicating renewed outburst activity, 24 years after the discovery outburst. The outburst had an amplitude of $\sim$ 9 magnitudes and lasted for several weeks, see Figure \ref{fig:ew}. 

We present the first time-resolved optical spectroscopy of GW Lib in (super)outburst (Section \ref{sec:outburst}) alongside time-resolved optical spectroscopy during quiescence before and after the outburst  (Section \ref{sec:calcium}). We present the possible system parameters in Section \ref{sec:systemparameters} and the discussion in Section \ref{sec:discussion}.

\section[]{Observations and reduction}
\label{sec:observations}

\begin{table*}
\begin{minipage}{2\columnwidth}
  \begin{center}
  \caption{Observations of GW Lib}
  \label{tab:observations}
  \begin{tabular}{l l c c c c c c c}
    \hline
    Telescope & Instrument &  Date & HJD &Exp time &$n$ spectra & Orbital & $\lambda$ - range & Seeing\\
    & &&  &    &  & coverage                           &       &        \\
    & & & (start) &(s) & & ($T_{\mathrm{obs}}/P_{\mathrm{orb}}$)& (\AA) & ($''$)\\ 

    \hline
    Magellan &B\&C  &06/07/2004&2453193.491&60 &62 &1.00 &3500 - 5100&0.5-1.0\footnote{From Magellan Guide Camera Seeing at \texttt{www.lco.cl}}\\
    &  & 07/07/2004&2453194.471& 60& 66 &1.02&&0.7-1.2 \\
    &     &&  &&& &&\\
    TT &FAST&14/04/2007&2454204.859 &60&1&--&3300 - 7600& 1-2\\
    &&14/04/2007&2454204.861&30&2&--&&1-2\\
    &&20/04/2007&2454210.859&15& 2&--&&-- \\
    &&09/05/2007&2454229.833&120& 1&--&&2\\
    &&09/05/2007&2454229.836&300& 1&--&&2\\
    &&11/05/2007& 2454231.817&120& 2&--&&2\\
    &&15/05/2007&2454235.813&480& 1&--&&--\\
    &&18/05/2007&2454238.894&600& 1&--&&--\\
    &&20/05/2007&2454240.811&480& 1&--&&--\\
    &&23/05/2007&2454243.787&720& 1&--&&--\\
    &&09/06/2007&2454260.739&720& 1&--&&--\\
    &&16/06/2007&2454267.774&1200& 1&--&&--\\
    &&24/06/2007&2454275.691&1200& 1&--&&--\\
    &&        &&  &  & &&\\
    TT &FAST&14/04/2007&2454204.886 &200&34& 1.86& 4150 - 5200&1-2\\
    &&15/04/2007&2454205.861&60&110&1.75&&1-2\\
    &&20/04/2007&2454210.882&60&70 &1.46&&1-2\\
    &&23/04/2007&2454213.858 &120&37 &1.68&&1-2\\
    &&10/05/2007&2454230.833 &300&18 &1.75&&2\\
    &&12/05/2007&2454232.833 &600&2& 0.16&&2\\
    &&17/05/2007&2454237.784 &300&22 &1.94&&1-2 \\
    &&        &  & & & &&\\
    Magellan &IMACS& 18/06/2007&2454270.531&30&73&1.09&3950 - 7100&0.6\\
    &      &&  &     &&&&\\
    WHT &ISIS& 24/07/2007 &2454306.365& 240&19& 1.08 &4200 - 5000/8050 - 8800&1\\
    & &25/07/2007 &2454307.350&240 &25 & 1.35&&1-2\\
    \hline
    \end{tabular}
    \end{center}
\end{minipage}
\end{table*} 
We briefly describe the facilities used below. Full details of the observations can be found in Table \ref{tab:observations}. 

\subsection{Telescopes and Instruments}
The telescopes used were the Magellan Telescopes operated by the Carnegie Institution of Washington at the Las Campanas Observatory in Chile, the Smithsonian Astrophysical Observatory's Tillinghast Telescope at the Fred Lawrence Whipple Observatory, located on Mount Hopkins near Amado, Arizona  and the William Herschel Telescope operated by the Isaac Newton Group on the island of La Palma. 

\subsubsection{Magellan Telescope}
Time-resolved optical spectroscopy was acquired during quiescence with the Baade 6.5-m telescope equipped with the Boller and Chivens Spectrograph (\textsc{b\&c}) on 2004 June 6 and 7.  The \textsc{b\&c} was used with a Marconi $2048\times515$ \textsc{ccd} with a 13.5 microns pixel size and a 1200 lines mm$^{-1}$ grating covering the spectral interval $3500 - 5100$\AA. A slit width of $0.8$~arcsec gave a dispersion of 0.79\AA~pixel$^{-1}$ and a resolution of 2.0\AA. 

On 2007 June 19  the telescope was equipped with the Inamori-Magellan Areal Camera and Spectrograph (\textsc{imacs}:  \citealt{bigelow2003}) to acquire time-resolved spectroscopy of GW Lib.  I\textsc{macs} was used with the f/4 camera and the long-slit-mask with a 0.7 arcsec slit width and a 600 lines mm$^{-1}$ grating centred at $5550$\AA. The spectra were dispersed along the short-axis of four SITe \textsc{ccd}s in the \textsc{imacs} mosaic detector. The \textsc{ccd} detectors were binned two by two during the observations. This instrumental setup provided a dispersion of $0.76$\AA~pixel$^{-1}$ and a spectral resolution of 1.29\AA\,\textsc{fwhm} in the spectral interval $3950-7100$\AA. 

\subsubsection{Tillinghast Telescope}
The 1.5-m reflector Tillinghast Telescope (\textsc{tt}) equipped with the \textsc{fast} instrument \citep{fabricant1998} acquired 7 epochs of time-resolved spectroscopy and 16 epochs of single spectroscopic frames in service mode. 
 For the time-resolved observations, the \textsc{fast} instrument was equipped with a 2688 $\times$ 512 UA STA520A \textsc{ccd} chip with a 15 microns pixel size. 
A grating of 1200 lines mm$^{-1}$ covering $4150 - 5200$\AA~and a 1.5~arcsec slit width was used to deliver a dispersion of 0.38\AA~pixel$^{-1}$ and spectral resolution of 0.86\AA\, \textsc{fwhm}. 

For the single frames the telescope was used in service mode with a grating of 300 lines mm$^{-1}$ and an aperture of 3.0~arcsec to cover the spectral interval $3300 - 7600$\AA. The pixels were binned two by two to give a dispersion of $1.47$\AA~pixel$^{-1}$ and a resolution of 5.9\AA. 

\subsubsection{William Herschel Telescope}
On two successive nights, three months after the 2007 outburst, we acquired time-resolved spectroscopic data  using the  4.2-m William Herschel Telescope (\textsc{wht}) in combination with the two-armed Intermediate dispersion Spectrograph and Imaging System (\textsc{isis}).
The blue arm of the spectrograph was equipped with a 4096 $\times$ 2048 EEV12 \textsc{ccd}.
The slit width used was 1.0~arcsec and used with the R1200B grating with 1200 lines mm$^{-1}$ providing a wavelength coverage of $4200 - 5000$\AA. This setup delivers a dispersion of 0.224\AA\, pixel$^{-1}$.
The red arm was equipped with a 4096 $\times$ 2048 Red+ \textsc{ccd} with 15 microns pixels. 
Using the R1200R grating with  1200 lines mm$^{-1}$ the unvingetted wavelength coverage was $8050 - 8800$\AA~with a dispersion of 0.243\AA~pixel$^{-1}$. The spectral resolution for both arms was 0.62\AA. 

\subsection{Reduction}
The different sets of spectroscopic data were reduced using several reduction packages.

\subsubsection{FAST service mode}
The single frame spectra obtained with the \textsc{fast} spectrograph in service mode were extracted using the spectral extraction pipeline provided by the Telescope Data Center at the Harvard-Smithsonian Center for Astrophysics. This tailored pipeline is based on the \textsc{iraf} \textsc{procd} package, see \citet{tokarz1997}.

\subsubsection{Magellan IMACS}
The \textsc{imacs} data were bias and flat-field corrected in the standard way using \textsc{iraf}. The spectra were extracted from each \textsc{ccd} frame with the \textsc{iraf} \textsc{kpnoslit} package. The pixel-to-wavelength calibration was derived from cubic spline fits to HeNeAr arcs acquired during the observations. Fitting 10 - 20 lines per frame gave a RMS $< 0.017$\AA~per chip.  

\subsubsection{All others}
All other data sets were reduced in the following way. The average bias and flat-field correction was carried out using the \textsc{figaro} package from \textsc{starlink} and nightly average bias and tungsten frames. \textsc{Pamela} was used for the optimal extraction of the spectra \citep{marsh1989}. 
Regular CuAr arc lamp exposures allowed us to establish an accurate wavelength scale for each spectrum through interpolation between the nearest arcs in time. Each arc frame has been fitted with a 5th order polynomial to 10-40 lines to give a typical RMS of 0.1 pixel. The individual spectra were normalised to the continuum level using a spline fit to selected continuum regions.

\section[]{GW Lib in Outburst}

\begin{figure}
  \includegraphics[width=\columnwidth]{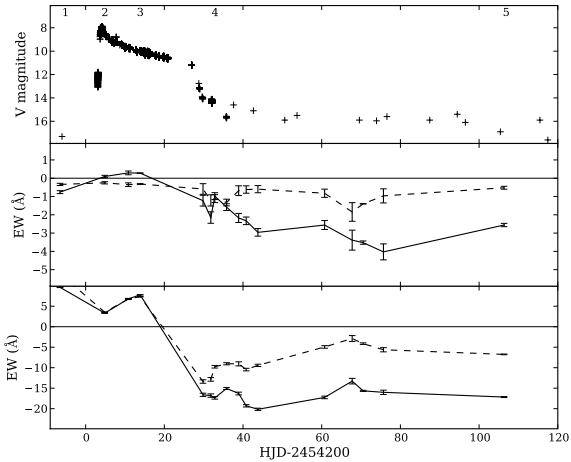}
  \caption{Time evolution of the equivalent widths of the principal spectral lines. The July 2004 quiescent epoch is plotted at an arbitrary negative HJD for  comparison.  In the bottom frame: \textit{solid} line for H$\beta$ and \textit{dashed} line for H$\gamma$. In the middle panel: \textit{solid} line for He~\textsc{i}~4921\AA, represents the general evolution of the He~\textsc{I} lines  and  \textit{dashed} line for He~\textsc{ii}~4685\AA. The top panel shows the V magnitude of the outburst, as provided by the \textsc{aasvo} where the numbers indicate position of the spectra from Figure \ref{fig:averagespectra} which are discussed epochs in Section \ref{subsection:spectralevolution}. 
}
  \label{fig:ew}
\end{figure}

\label{sec:outburst}
\begin{figure*}
\begin{minipage}{2\columnwidth}
  \includegraphics[width=\columnwidth]{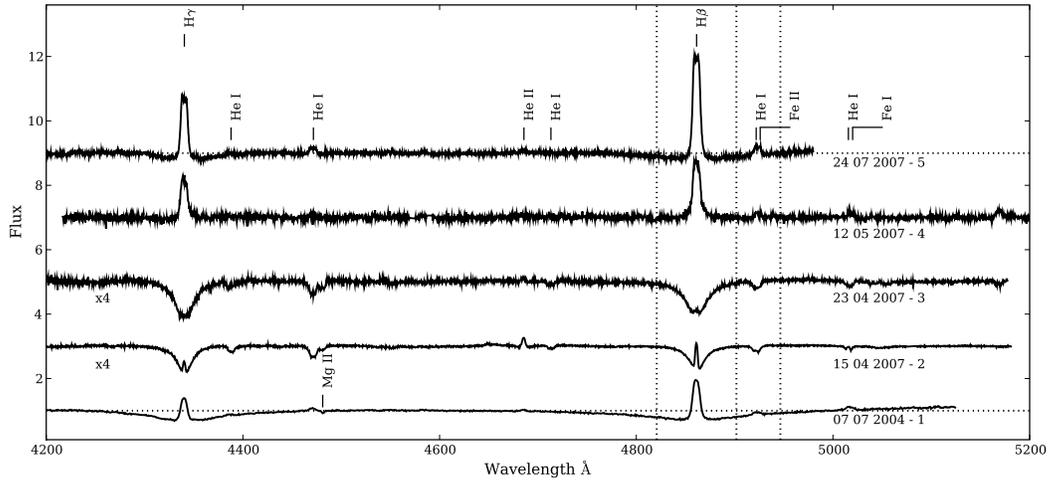}
  \caption{Comparison of average representative normalised spectra of GW~Lib sampling the spectral evolution before and during the 2007 super-outburst. The April 2007 spectra have been multiplied by a factor of 4 to magnify the spectral features and allow for a better comparison. The vertical \textit{dotted} lines annotate the $\pm 2500$ km s$^{-1}$ window for the Balmer lines and the $\pm 1500$ km s$^{-1}$ band for helium lines while the horizontal \textit{dotted} lines the normalised continuum flux.}
  \label{fig:averagespectra}
\end{minipage}
\end{figure*}

The April 2007 outburst of GW Lib triggered our target of opportunity program at the \textsc{tt}.  Both time-resolved spectroscopy and single frame observations of GW Lib were obtained during this period of activity. See Table \ref{tab:observations} for specific details.

\subsection{Spectral evolution}
\label{subsection:spectralevolution}
In Figure \ref{fig:averagespectra} average representative normalised spectra during different stages of the outburst of GW Lib are presented. The bottom spectrum is during quiescence 3 years before the 2007 outburst, the second spectrum is a day after outburst peak, the third during slow decline, the fourth during rapid decline and the fifth during the final decline towards quiescence. For reference, the top frame in Figure \ref{fig:ew} shows the outburst light curve in the V-band, as provided by \textsc{aavso} and its network of observers. The dates corresponding to the spectra plotted in Figure \ref{fig:averagespectra} are marked. The pre-outburst quiescent spectrum is given an arbitrary negative HJD for plotting purposes.  

The single frame spectra at different nights throughout the outburst make it possible to trace the evolution of various spectral features. 
We have measured the equivalent widths (\textsc{ew}) of the lines as a function of the outburst. For the Balmer lines, the overall continuum is normalised but any underlying absorption within the line was not removed. This results in positive (\textsc{ew}) values during absorption-dominated epochs and negative values otherwise. For consistency, the window for these lines is $\pm 2500$ km~s$^{-1}$ for all epochs, which is limited by the proximity of the helium lines. For the He lines, a slightly narrower window of $\pm 1500$ km~s$^{-1}$ gives a consistent and comparable \textsc{ew} measurement, avoiding neighbouring lines. The spectra are locally normalised to avoid any contamination from underlying absorption due to nearby Balmer lines.  
Both regions are marked in Figure \ref{fig:averagespectra} together with the continuum level. The bottom panel in Figure \ref{fig:ew} shows the \textsc{ew} evolution for the Balmer lines and the middle panel the same for the helium lines. The actual values for all lines and at all observed epochs are given in Table \ref{tab:equivalentwidth}. 
We note that the spectral lines display rather complex changes as different components come and go. A single (\textsc{ew}) value clearly does not fully capture these details. Therefore, to complement the measured (\textsc{ew}), a more detailed analysis of the separate components within the line profiles is provided below. 

\paragraph*{2004 July 7 - 1}
During deep quiescence, 3 years before outburst, the average spectrum of GW Lib was typical for a CV spectrum with a low accretion rate. The accretion disc has very narrow Balmer emission lines which show a dip in their centres reflecting an underlying double-peaked profile (not visible on the scale of Figure \ref{fig:averagespectra}). This profile shape is expected for optically thin accretion discs viewed at a low inclination angle \textit{i}. 
The H$\beta$ profile is fitted with two Gaussians simultaneously to the absorption trough and the narrow emission line component. The latter gives a \textsc{fwhm} of $7.82 \pm 0.56$\AA\, corresponding to $\sim 482$ km~s$^{-1}$ (all errors quoted in this paper are $3\sigma$ errors).
Absorption troughs flank the emission features with a \textsc{fwzi} of $\sim 20,000$ km~s$^{-1}$, and can only originate from the WD as these velocities are too high for Kepler velocities in the accretion disc. This indicates a small accretion luminosity and a very low mass transfer rate such that the primary WD is visible. The measured \textsc{ew} of the Balmer lines are dominated by the WD absorption. The He~\textsc{i} lines are all in emission. A small, but significant, amount of He~\textsc{ii} is present in emission with an \textsc{ew} of $-0.34\pm 0.18$\AA\, in the average spectrum. Since the disc and companion star are expected to be too cool during quiescence to produce much He~\textsc{ii} emission, this line emission may be due to the relatively hot WD ionising gas near it. The feature is unfortunately too weak to permit any time-resolved analysis.  We also note that an absorption line at Mg~\textsc{ii} 4481.15\AA~ is present with an \textsc{ew} of $0.24\pm 0.09$\AA\, measured in a $\pm 300$~km~s$^{-1}$ window and a \textsc{fwhm} of $3.18 \pm 0.35$\AA. This is likely formed in the photosphere of the WD and such metal lines may be expected in accreting systems were metal-rich gas is deposited onto the WD surface. The determination of the gravitational redshift of such lines could in principle provide a direct measurement of the mass of the WD. However, our resolution prohibits a reliable measurement of the velocity of this weak feature as it is blended with the nearby He~\textsc{i} line.

\paragraph*{2007 April 15 - 2}
Two days after the rise into outburst started, the average spectrum of GW Lib shows broad absorption troughs in the Balmer lines. Given the 9 magnitude increase due to accretion these absorption features can no longer be associated with the WD but suggest an origin in the now optically thick accretion disc, supported by a \textsc{fwhm} of $23.05\pm0.36$\AA. The lines also show strong emission peaks, see spectrum 2 in  Figure \ref{fig:averagespectra}. This is also seen in the H$\alpha$ emission profile obtained by \citet{hiroi2009} around this time. With a \textsc{fwhm} of only $3.14 \pm 0.66$\AA, corresponding to $\sim 194$ km~s$^{-1}$, the peak in H$\beta$ has less than half the width of the emission during quiescence. 
The He~\textsc{i} lines are all in absorption, with a hint of a very narrow emission peak at the centre, while the He~\textsc{ii} line at 4685\AA\, is in emission at \textsc{ew} = $-0.25 \pm 0.19$\AA. Finally Fe is marginally detected in absorption. We will revisit the discussion on the origin of these lines in section \ref{subsec:timeresolvedspectrum}. 

\paragraph*{2007 April 23 - 3}
Ten days after the beginning of the outburst, GW Lib slowly started to fade and the spectrum evolved. The strong and narrow emission in the H$\beta$ and H$\gamma$ lines decreased in strength, increasing the overall \textsc{ew} slightly. The \textsc{ew} of the He~\textsc{i} lines increases similarly suggesting the same narrow emission was indeed present.  The accretion disc is still visible as an optically thick disc in absorption in the Balmer lines with a \textsc{fwhm} of $30.46 \pm 0.26$\AA. The He~\textsc{ii} emission decreases slightly to an \textsc{ew} of $-0.32\pm0.06$\AA.  

\paragraph*{2007 May 12 - 4}
A month after outburst, as the system dropped from the outburst plateau (Figure \ref{fig:ew}) and the luminosity decreased rapidly, the spectrum had changed drastically, see spectrum 4 in Figure \ref{fig:averagespectra}. The accretion disc contribution to the lines changed from an optically thick flow back into a chiefly optically thin disc, producing shallow double-peaked emission profiles in the Balmer line without the high velocity absorption, leading to a negative \textsc{ew}. The emission component in H$\beta$ is visible again and now has a \textsc{fwhm} of  $8.59 \pm 0.03$\AA, close to the value of the pre-outburst emission line. As can be seen in Figure \ref{fig:ew}, the \textsc{ew}s of all lines make a small jump. WZ~Sge type DNe often show brief brightenings shortly after their main outburst, referred to as echo-outbursts \citep{patterson2002}. It remains unclear due to the sparse light curve sampling whether the observed behaviour in the lines is related to similar events.

\begin{figure}
  \includegraphics[width=\columnwidth]{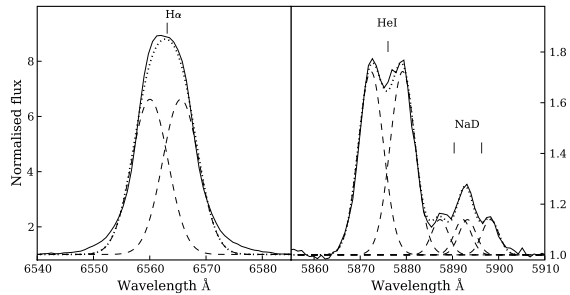}
  \caption{Normalised emission line profiles of H$\alpha$ (left) and the He~\textsc{i} - Sodium doublet (right). The \textit{solid} line represents in both cases the average profile over 1 orbital period. The \textit{dashed} lines are the various Gaussian profiles fits and the \textit{dotted} line is the sum of all single Gaussians. As is clear from these fits, the Sodium doublet has, most likely, a very similar origin as the H$\alpha$ and the He~\textsc{i} lines. }
  \label{fig:sodiumprofile}
\end{figure}

\paragraph*{2007 June 19}
The Magellan spectra from June 2007 resemble the \textsc{wht} July spectra in shape, see top spectrum in Figure \ref{fig:averagespectra} for the average spectrum. In addition, these spectra also cover the H$\alpha$ and the He~\textsc{i} 5876.62\AA\,  line. An interesting feature next to the strong He~\textsc{i}  line is the Sodium doublet (Na~D 5889.95\AA -5895.92\AA) seen in emission, see Figure \ref{fig:sodiumprofile}. Due to the close proximity of GW Lib, interstellar absorption is expected to be weak, and indeed no evidence for diffuse interstellar band (\textsc{dib}) absorption features can be found in the spectrum. Instead, the doublet is dominated by broad emission components.
To strengthen this case, the Sodium doublet is also present in emission at 6154.23\AA-6160.76\AA. In the orbital average spectrum, the feature can be fitted with two double Gaussians (plotted in Figure \ref{fig:sodiumprofile}) with the peak separation set to $250$~km~s$^{-1}$. The H$\alpha$ line has the same separation where for the nearby He~\textsc{i} the line separation is $350$~km~s$^{-1}$. The offset of the centre of the double-peaked profiles from the rest wavelength for all lines is, within errors, $-10 \pm 5$~km~s$^{-1}$. These similar fitting values suggest a shared origin in the accretion flow for all lines. Time-resolved analyses of the He~\textsc{i} and the Balmer lines give similar values compared to those in the July 2007 epoch  (see section \ref{subs:emission} and Table \ref{tab:cavelocity}).

\paragraph*{2007 July 24 - 5}
Three months after outburst, during the slow decline towards quiescence, the system was still 1-2 mags brighter than the pre-outburst magnitude, and the double-peaked profile from the disc proved stronger against the dropping continuum compared to the previous phase (compare spectrum 4 to spectrum 5 in Figure \ref{fig:averagespectra}). The system has cooled sufficiently such that the WD absorption again flanks the emission lines in the Balmer lines, decreasing the \textsc{ew} of H$\gamma$.  Since both the disc and the WD are expected to be slightly hotter and brighter compared to the pre-outburst quiescence state it does not yet resemble spectrum 1 in Figure \ref{fig:averagespectra}. For comparison, the WD in WZ~Sge was still cooling after 17 months of post-outburst observations, see \citet{long2004}.
The He~\textsc{i} lines are all in emission and the He~\textsc{ii} emission line has an \textsc{ew} of $-0.52 \pm 0.24$\AA. The \textsc{fwhm} of the emission component in H$\beta$ is $7.65 \pm 0.03$\AA~and has thus reached the value observed in the pre-outburst epoch even though the emission line flux itself is still much larger.  

\begin{table*}
\begin{minipage}{2\columnwidth}
\begin{center}
  \caption{Equivalent widths in \AA\, $\pm 3\sigma$ error, of various emission lines as function of the outburst. The Balmer lines are measured in a $\pm 2500$~km~s$^{-1}$ window, where the helium lines are measured in a $\pm 1500$~km~s$^{-1}$ window }
  \label{tab:equivalentwidth}
  \begin{tabular}{l c c c c c c c}
    \hline
Epoch & HJD & H$\beta$ & H$\gamma $ & HeI 4387 & HeI 4471& HeII 4685 & HeI 4921\\
&(start) & & & & & & \\
\hline
1& 2453194.471 & 9.61 $\pm$ 0.18 & 12.93 $\pm$ 0.15 & -0.29 $\pm$ 0.18 & -0.74 $\pm$ 0.12 & -0.34 $\pm$ 0.18 & -0.76 $\pm$ 0.24 \\
2& 2454204.859 & 3.32 $\pm$ 0.16 & 3.55 $\pm$ 0.17 & 0.04 $\pm$ 0.18 & 0.47 $\pm$ 0.18 & -0.25 $\pm$ 0.19 & 0.10 $\pm$ 0.19 \\
& 2454210.859 & 6.79 $\pm$ 0.37 & 6.75 $\pm$ 0.37 & 0.04 $\pm$ 0.30 & 1.19 $\pm$ 0.30 & -0.35 $\pm$ 0.30 & 0.29 $\pm$ 0.31 \\
3& 2454213.858 & 7.78 $\pm$ 0.07 & 7.32 $\pm$ 0.09 & 0.50 $\pm$ 0.06 & 1.03 $\pm$ 0.07 & -0.32 $\pm$ 0.06 & 0.28 $\pm$ 0.03 \\
& 2454229.833 & -16.61 $\pm$ 1.19 & -13.37 $\pm$ 1.26 & 0.29 $\pm$ 0.91 & -2.84 $\pm$ 0.97 & -0.59 $\pm$ 0.87 & -1.23 $\pm$ 0.89 \\
& 2454231.817 & -16.80 $\pm$ 1.23 & -12.85 $\pm$ 1.33 & -0.00 $\pm$ 0.98 & -1.30 $\pm$ 0.98 & -1.21 $\pm$ 0.92 & -2.15 $\pm$ 0.94 \\
4& 2454232.833 & -17.37 $\pm$ 0.90 & -9.77 $\pm$ 0.95 & -0.96 $\pm$ 0.62 & -0.01 $\pm$ 0.62 & -1.11 $\pm$ 0.65 & -0.98 $\pm$ 0.56 \\
& 2454235.813 & -15.07 $\pm$ 0.74 & -9.04 $\pm$ 0.78 & -0.36 $\pm$ 0.58 & -0.84 $\pm$ 0.57 & -1.33 $\pm$ 0.55 & -1.58 $\pm$ 0.56 \\
& 2454238.894 & -16.30 $\pm$ 1.01 & -9.06 $\pm$ 1.48 & -0.83 $\pm$ 1.07 & -1.06 $\pm$ 0.95 & -0.68 $\pm$ 0.80 & -2.17 $\pm$ 0.74 \\
& 2454240.811 & -19.30 $\pm$ 0.84 & -10.46 $\pm$ 0.94 & -1.15 $\pm$ 0.71 & -1.43 $\pm$ 0.66 & -0.62 $\pm$ 0.61 & -2.33 $\pm$ 0.62 \\
& 2454243.787 & -20.15 $\pm$ 0.82 & -9.43 $\pm$ 0.82 & -1.51 $\pm$ 0.64 & -1.32 $\pm$ 0.62 & -0.60 $\pm$ 0.59 & -2.96 $\pm$ 0.62 \\
& 2454260.739 & -17.25 $\pm$ 0.95 & -4.94 $\pm$ 0.93 & -1.09 $\pm$ 0.74 & -0.29 $\pm$ 0.68 & -0.82 $\pm$ 0.68 & -2.56 $\pm$ 0.73 \\
& 2454267.743 & -13.28 $\pm$ 2.05 & -2.88 $\pm$ 2.15 & -0.57 $\pm$ 1.67 & -0.74 $\pm$ 1.54 & -1.84 $\pm$ 1.53 & -3.39 $\pm$ 1.64 \\
& 2454270.531 & -15.65 $\pm$ 0.41 & -4.14 $\pm$ 0.57 & -0.65 $\pm$ 0.46 & -2.04 $\pm$ 0.42 & -1.40 $\pm$ 0.30 & -3.52 $\pm$ 0.27 \\
& 2454275.691 & -16.01 $\pm$ 1.58 & -5.62 $\pm$ 1.54 & -1.32 $\pm$ 1.28 & -3.62 $\pm$ 1.26 & -0.97 $\pm$ 1.16 & -4.03 $\pm$ 1.31 \\
5& 2454306.355 & -17.16 $\pm$ 0.35 & -6.76 $\pm$ 0.27 & -0.75 $\pm$ 0.24 & -1.47 $\pm$ 0.21 & -0.52 $\pm$ 0.24 & -2.56 $\pm$ 0.27 \\
\hline
 \end{tabular}
 \end{center}
\end{minipage}
\end{table*}

\subsection{Time-resolved spectra}
\label{subsec:timeresolvedspectrum}
Several epochs (see Table \ref{tab:observations}) of time-resolved spectroscopy allows us to characterise the line behaviour in more detail.  During outburst, these time-resolved spectra showed considerable orbital dependence in the Balmer, He~\textsc{i} and He~\textsc{ii} line profiles. We binned the spectra in bins of 1/20th of the orbital period using $P = 0.05332 \pm 0.00002$~days \citep{thorstensen2002}. The Balmer profiles were fitted with a double Gaussian with the peak, \textsc{fwhm} and the offset from the rest wavelength as free parameters. The He~\textsc{i} and He~\textsc{ii} profiles are fitted with a single Gaussian. All fits show that the peak and \textsc{fwhm} of the line components are constant within 10 per cent of their average values and as these particular parameters show no sign of significant orbital dependence, they were thus fixed to their mean values for further analysis, keeping their velocity offsets as free parameters. In addition to our profile fits, all results were checked against a traditional double Gaussian a\-na\-ly\-sis (\citealt{schneider1980}) in combination with a diagnostic diagram.  

The radial velocity curves of our line components were fitted with the function $V(\phi) = \gamma - K\sin(2\pi\phi - 2 \pi \phi_0)$ and individual velocity data points were weighted according to their errors as derived from the profile fits. These errors were scaled such that the radial velocity curve fit has a goodness of fit close to 1.
Here $\gamma$ is the systemic velocity, $K$ the radial velocity semi-amplitude of the absorption and/or emission line and  $\phi_0$ the phase offset relative to the ephemeris.  The ephemeris zero point will be derived later in this paper (section \ref{sec:calcium}) and is used throughout this paper when calculating orbital phases. Due to the lack of precision in the binary orbital period, the ephemeris cannot be extrapolated very far in time and thus the binary phase is arbitrary except for the July 2007 epoch where the zero-point is measured.

\begin{figure}
  \includegraphics[height=70mm]{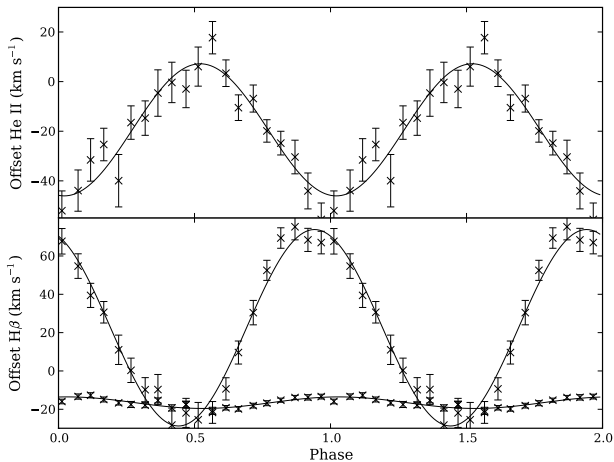}
  \caption{Top panel shows the radial velocity curve of the He~\textsc{ii} emission. The radial velocity curves of the H$\beta$ profile (bottom) are fitted with a low amplitude emission component and a high amplitude absorption component. Both are from the phase-binned 2007 April 15 data, showing two orbital cycles.}
  \label{fig:velocity}
\end{figure}

We discuss the results of the fits for the H$\beta$ and He~\textsc{ii} profiles using the 2007 April 15 data as this set has the highest signal to noise, the best time resolution and the most features present for analysis. The behaviour of the lines during the other nearby epochs were the same, albeit determined with lower precision. The results can be found in Figure \ref{fig:velocity} and Table \ref{tab:velocity}. Both H$\beta$ and H$\gamma$ have similar amplitude and phase fits. The deep and broad absorption troughs can be identified with the accretion disc. However, they do not trace the WD movement as the velocity semi-amplitude is too large at $51.3 \pm 6.5$~km~s$^{-1}$. The radial velocity of the primary WD is likely much smaller. Interestingly, this amplitude is consistent with that of the double-peaked emission profile during quiescence. On 2004 July 6, the velocity curve fit to H$\beta$ emission gives a semi-amplitude of $45.1 \pm 7.3$~km~s$^{-1}$. This similarity suggests that the distortions in the disc leading to the observed radial velocity curve of both the disc emission during quiescence and the broad disc absorption during outburst are rather similar. Unfortunately, due to the uncertainty in the binary ephemeris, we cannot compare their relative phasing. Similar amplitudes, systemic velocities and phase zeros were found for the He~\textsc{i} lines, suggesting that their profiles trace the same components. In contrast, the He~\textsc{ii} line moves in anti-phase and has half the amplitude. 
This highlights the difficulty with associating either of these radial velocity amplitudes with that of the primary WD.

In both the H$\beta$ and H$\gamma$ profile, a second Gaussian is fitted to the narrow emission components in their cores. This component shows a remarkably small velocity amplitude of $3.0 \pm 1.6$~km~s$^{-1}$ (see Table \ref{tab:velocity}). At this epoch, the arc fit is good to 0.003\AA~at the position of H$\beta$ which is smaller than 1/100th of the 0.38\AA/pixel scale. 1$\sigma$ errors on individual velocity measurements are much larger ($\sim$0.085 pixels) than this and are thus dominated by the statistical error. These errors are then propagated to provide the formal errors on the fit parameters as listed thus suggesting that this small amplitude is measured to a significant degree.

This strong central emission component is present in those epochs close to the start of the outburst in the Balmer lines and a hint is present in the He~\textsc{i}lines. On 2007 April 20 the dynamics of this component is similar to behaviour on April 15. By 2007 April 23 the emission component has faded into the absorption lines, which maintain their radial velocity semi-amplitude throughout. The systemic velocity of this narrow emission component is consistent with that of the He~\textsc{ii} emission, while the broad absorption components in the Balmer lines have a rather different $\gamma$ which is perhaps due to a slowly precessing disc.

\begin{table*}
\begin{minipage}{120mm}
  \caption{Velocity profile parameters on 2007 April 15}
  \label{tab:velocity}
  \begin{tabular}{l c c c c}
    \hline
    Line & $\gamma$ & $K$ & $\phi_0$ & Identification \\
         &  $\mathrm{km\,s}^{-1}\pm 3 \sigma$ & $\mathrm{km\,s}^{-1} \pm 3 \sigma$ &  $\pm 3 \sigma$&\\
    \hline
    H$\beta$ emission & $-16.6 \pm 0.8 $ & $3.0  \pm 1.6 $ & $ 0.28 \pm 0.06$ & Slingshot/WD? \\
    H$\beta$ absorption & $22.5 \pm 4.4 $ & $51.3  \pm 6.5$ & $0.19 \pm 0.02$ & Accretion disc \\
    H$\gamma$ emission & -- & -- &  --& --\\
    H$\gamma$ absorption & $17.0  \pm 4.8$& $ 43.6 \pm 6.8$& $0.18  \pm 0.02$& Accretion disc \\
    He~\textsc{ii} emission & $-19.4  \pm 4.7$& $26.6  \pm 6.8 $& $0.77 \pm 0.04$& -- \\
    \hline
  \end{tabular}
\end{minipage}
\end{table*}

GW~Lib is one of the few CVs that have been followed through the outburst spectroscopically.  
For comparison, the spectral evolution of SS~Cyg (figure 5 in \citealt{hessman1984}) showed broad, double-peaked emission lines that are gradually being overtaken by the rising continuum, followed by the formation of line wings in absorption.  This qualitative behaviour is expected as the accretion disc makes a rapid change from a low $\dot{M}$ optically thin configuration to a high $\dot{M}$ optically thick flow at the onset of the outburst. Near maximum light, SS~Cyg  shows on top of the absorption from the accretion disc, narrow emission cores which decay again after outburst. These are very similar to the spectral changes presented here for GW~Lib, though no dynamical properties are available to compare the components quantitatively. 
Unfortunately, the very low binary inclination of GW Lib makes it difficult to study the accretion disc dynamics during the main outburst period. The absorption dominated disc lines are unsuitable for Doppler tomography and we therefore cannot search for disc asymmetries such as the tidal spirals seen in other DNe (e.g. IP~Peg: \citealt{steeghs1997}; U~Gem: \citealt{groot2001}). When disc emission returns, little structure can be detected in its marginally double-peaked lines.

The stationary emission component in the spectra, visible only during the first couple of days of the outburst, does not obviously fit in with the typical components expected to dominate the line emission in a mass transferring binary system. The low velocity suggests a location near the centre of mass in the orbital plane or along the axis through the centre of mass, perpendicular to the orbital plane. GW ~Lib is not the first DN to show these features. Several DN systems have been reported to show low amplitude strong emission components in the Balmer lines.

\citet{steeghs1996} report low velocity emission components in the H$\alpha$ and He~\textsc{ii} 4686\AA\, lines of the DNe IP~Peg and SS~Cyg and proposed slingshot prominences from the donor star as a possible origin for these features since their known system parameters rule out an origin near either the WD or the surface of the donor star. The rapidly co-rotating donor star may form prominences and such magnetic loops would be pulled towards the WD and could potentially find an equilibrium in between the L1 point and the WD. As the prominence material is illuminated by the disc during outburst, it becomes visible as an emission source which is co-rotating with the binary orbit but located in the region near the centre of mass and thus displays very little radial velocity. If we follow their recipe, the observed H$\beta$/He~\textsc{ii} ratio is consistent with a gas temperature in the 10,000 - 15,000~K range. 

In GW Lib, an alternative origin of the low-velocity emission may be near the surface of the WD, if its orbital velocity ($K_1$) is very small. Low state AM~CVn systems often show such narrow emission components (\citealt{morales-rueda2003}, \citealt{roelofs2006}).  The low mass ratio derived from the late superhump by \citet{kato2008} together with the low orbital inclination would indeed imply a very low value for $K_1$ in GW Lib. We will revisit these interpretations and the implications for the system parameters in section \ref{sec:systemparameters}.

\section[]{GW Lib post outburst}
\label{sec:calcium}

\begin{figure}
  \includegraphics[height=70mm]{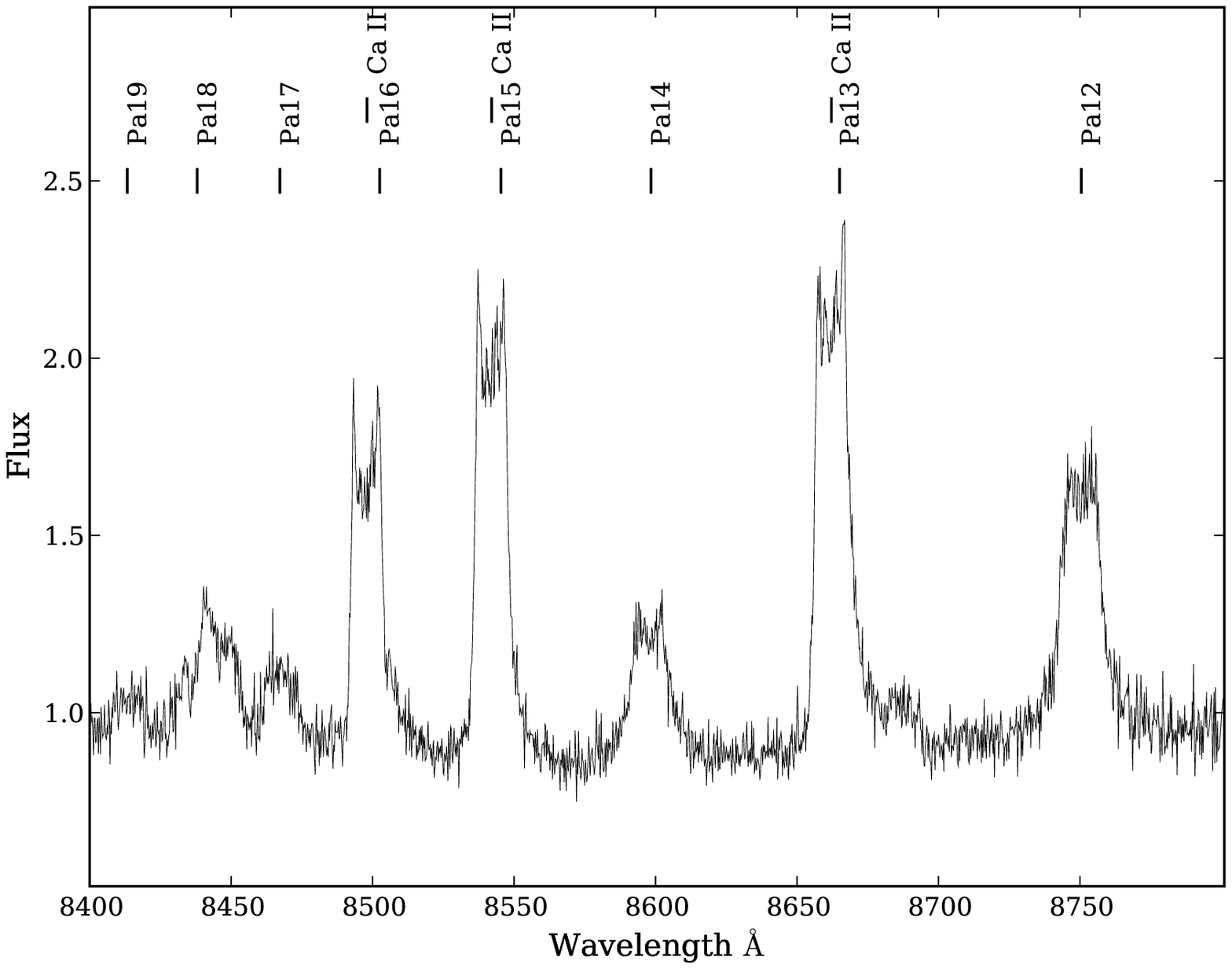}
  \caption{The average normalised I-band spectrum of GW Lib on July 24, 2007 showing the Paschen series and the double-peaked Ca~\textsc{ii} lines. }
  \label{fig:calciumspectrum}
\end{figure}

Binary evolution tells us that the majority of the current CV population should have evolved towards short orbital periods with the mass donor star depleted to a low mass degenerate brown dwarf (e.g. \citealt{knigge2006}, \citealt{kolb1992}, \citealt{rappaport1983}).
Not only has this graveyard of CVs been elusive for many years (see \citealt{gaensicke2009}), our sample of short period CVs with reliable binary parameters, other than the orbital period, is very small. This is mainly due to the fact that despite the low mass transfer rates, the donor stars are extremely faint and it is very difficult to perform traditional radial velocity studies using photospheric absorption lines from the donor.
Most studies are confined to using the strong Balmer emission lines to try and constrain system parameters (e.g. table 2 in \citealt{marsh2001}, \citealt{john2007}, \citealt{neilsen2008}, \citealt{mennickent2006},  \citealt{szkody2000}).
The Ca~\textsc{ii} triplet at $8498.03$\AA, $8542.09$\AA\, and $8662.14$\AA\footnote{From The Atomic Line List Version 2.04: \texttt{http:/www.pa.uky.edu/\~ {}peter/atomic/}} offers advantages that more than make up for its relative weakness compared to the Balmer lines. It has a lower ionisation energy than either hydrogen and helium and is thus capable of being excited even by cool sources of radiation, and its thermal width and pressure broadening are much smaller than for hydrogen leading to sharper, more easily detected spikes of emission \citep{marsh1997}. Studies have shown that Ca~\textsc{ii} is also more accessible than the Balmer lines for the study of velocity gradients and turbulence. This is because the thermal velocity broadening is smaller in general and the Keplerian shear broadening starts to dominate at higher inclinations compared to the hydrogen lines \citep{horne1995}. These studies show a promising, but so far neglected avenue for emission line studies of CVs. The average normalised spectrum of GW~Lib in the I-band obtained on 2007 July 24 is plotted in Figure \ref{fig:calciumspectrum} and shows a strong Ca~\textsc{ii} triplet in emission.

\subsection{Emission from the secondary}

\begin{figure*}
\begin{minipage}{150mm}
  \includegraphics[width=150mm]{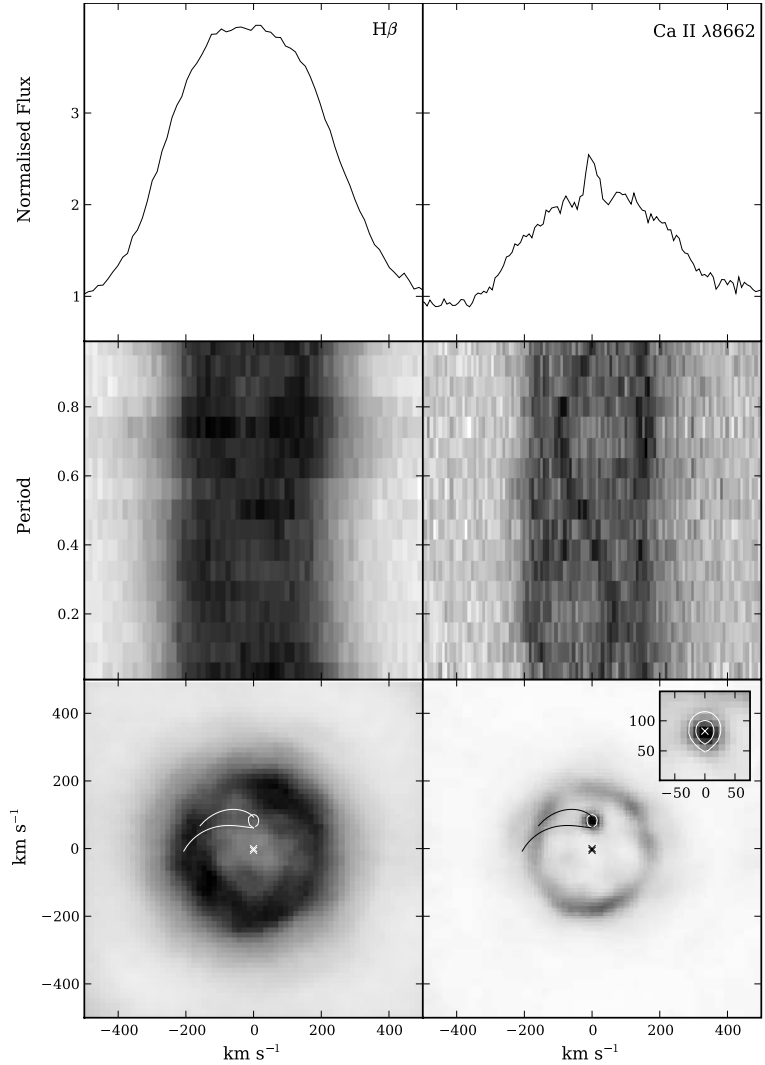}
   \caption{Comparison of the dynamics of the Balmer lines (H$\beta$, left column) versus Ca~\textsc{ii} (8662\AA, right column). The top row show average profiles in the reference frame of the donor star S-wave. The sharp donor component stands out in Ca~\textsc{ii} but the H$\beta$ profile is smooth. The middle row shows the phase-binned trailed spectrograms. The S-wave can be traced in between the disc peaks in Ca~\textsc{ii} only. The bottom row plots the Doppler tomograms showing the superior sensitivity and sharpness in the Ca~\textsc{ii} lines. The S-wave is mapped to a sharp spot consistent with donor star emission. Overlaying the Doppler tomograms are the positions of the WD, the Roche Lobe of the donor star and the streams for a model with $q=0.062$. The inset in the Ca~\textsc{II} Doppler map shows the Roche Lobe on the Donor star in more detail for both a $q=0.062$ and $q=0.23$ solutions.}
   \label{fig:trailer}
\end{minipage}
\end{figure*}

In Figure \ref{fig:trailer} (middle panels), we compare trailed spectrograms of H$\beta$ (left) and Ca~\textsc{ii} 8662\AA\, (right) as obtained 3 months after outburst. Whereas the Balmer line shows a shallow double-peaked profile and a rather blurry trail, the Ca~\textsc{ii} trail is much sharper even at a lower S/N. It shows a clear double-peaked profile from the accretion disc and an S-wave moving in between. No such S-wave appears to be present in the Balmer lines.

To highlight this component, we averaged the profiles in a frame co-moving with the S-wave and show these in the top panels of Figure \ref{fig:trailer}. The double-peaked profile from the accretion disc in this frame is averaged out, but the narrow S-wave is clearly visible on top on the broad profile in the Ca~\textsc{ii} case whereas its absence is noted in the Balmer profile. 

The S-wave is sharp and its amplitude is smaller than the velocity offset of the accretion disc peaks, suggesting it may originate from the secondary star rather than from the interaction point between the infalling stream and the accretion disc, the hotspot, since the latter would have a velocity equal to or larger than the outer disc edge.  

\begin{figure}
  \includegraphics[height=70mm]{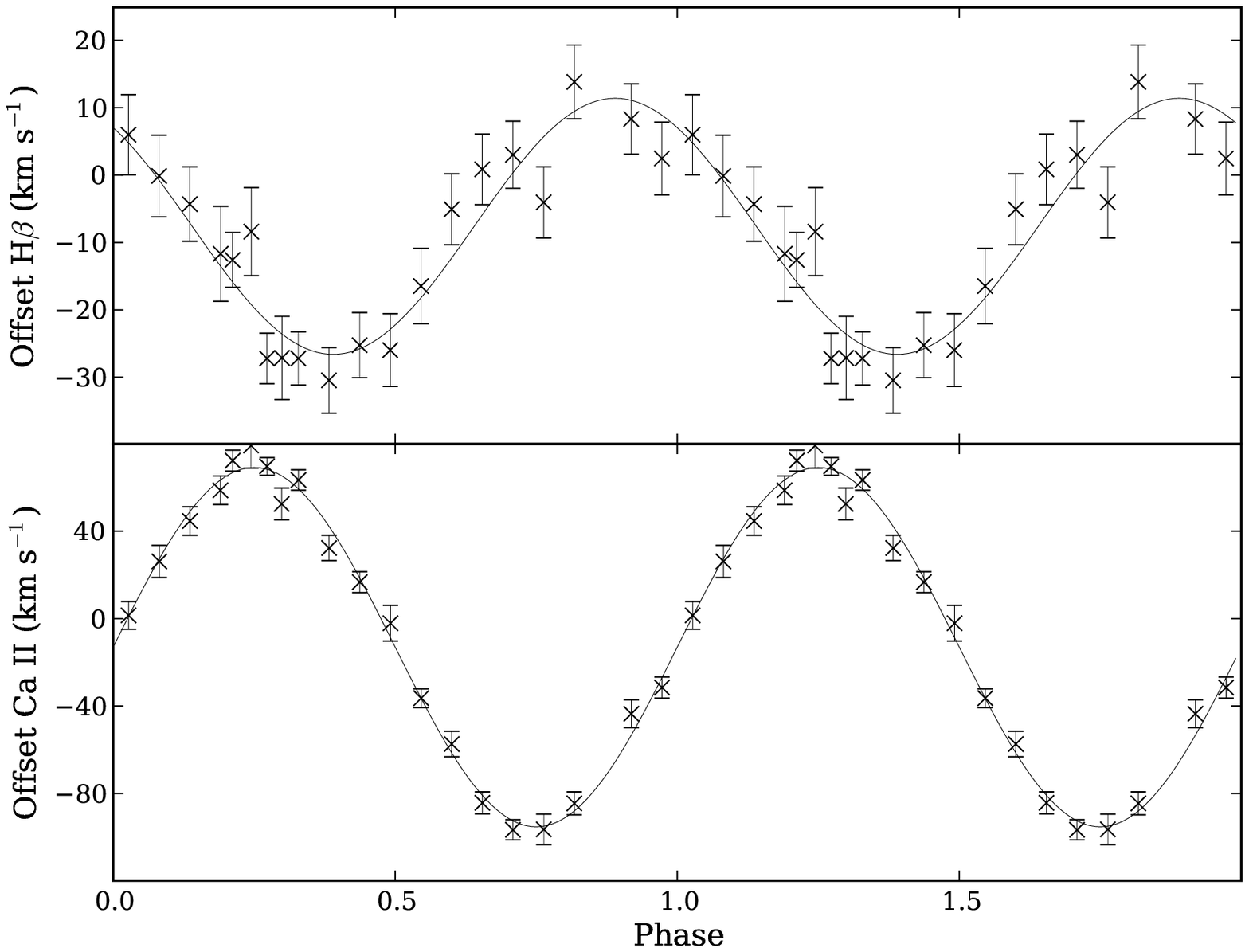}
  \caption{Top panel shows the radial velocity curve of the H$\beta$ emission. The bottom panel shows rotational velocity curve from the Ca~\textsc{ii} 8662\AA\,  S-wave emission component. Both are from the phase-binned 2007 June 24 data, showing two orbital cycles.}
  \label{fig:cavelocity}
\end{figure}

Assuming the emission does indeed arise from the surface of the donor star, the ephemeris can be determined as the crossing from blue-to-red of its velocity curve, which corresponds to inferior conjunction. This gives: 
\[
\mathrm{HJD} = 2454307.36867 + 0.05332 E
\]
This is used throughout the paper to define the orbital phase although the accuracy of the orbital period is not enough to give the outburst data discussed in Section 3 a definite orbital phase since the accumulated uncertainty is too large (Figure \ref{fig:cavelocity}). 

The radial velocity semi-amplitude of the emission is determined using two different methods. The first method is the fitting of a Gaussian to the S-wave line component followed by a radial velocity curve fit. The second method  is the localisation of the emission peak in the velocity-velocity plane using Doppler tomography.

For the first method, the spectra are phase-binned and a triple Gaussian profile with as variables the offsets, the peak heights and the \textsc{fwhm}s of the profiles is fitted where two Gaussians trace the disc features and the third is for the donor S-wave.  Both the peak height and the \textsc{fwhm} show a chaotic variability with a maximum amplitude of 8 per cent from the average value and were therefore fixed to their mean values, leaving just the velocity offset as fit parameter.  The offset velocity for the S-wave emission is plotted in Figure \ref{fig:cavelocity} together with a weighted radial velocity curve fit. For improved accuracy in the Ca~\textsc{ii} triplet, we performed a joint fit to the 3 triplet lines with a common velocity offset. We then find a semi-amplitude of $K_{\mathrm{Ca~\textsc{ii}}} = 82.2 \pm 4.9$~km~s$^{-1}$ for the S-wave in the Ca~\textsc{ii} profiles, see also Table \ref{tab:cavelocity}.

To complement our radial velocity fits, Doppler maps are created with a Maximum Entropy Method (\textsc{mem}; \citealt{marsh2001}) for both the Balmer and the Ca~\textsc{ii} $8662$\AA~lines, see bottom panels in Figure \ref{fig:trailer} for \textsc{mem} map of the time-resolved spectra on 2007 June 24. 
Given the fact that the lines from the Paschen series overlap with the Ca~\textsc{ii} triplet lines, we always ensured that the Doppler reconstructions fitted to both lines simultaneously. The underlying Paschen line distribution is rather featureless, and the sharp S-wave component is only present in the Ca~\textsc{ii} reconstructions. Even if such a Paschen contribution is not included, the Ca~\textsc{ii} tomograms show very similar features.  

The Balmer map (bottom left panel) shows a diffuse ring of emission from the accretion disc. Due to the sharper line profile, the Ca~\textsc{ii} map (bottom left panel) shows a much better defined accretion disc ring with a significant asymmetry at phases $\sim 0.10$, $\sim0.35$ and $\sim 0.70$. Furthermore, the S-wave component manifests itself as a sharp emission spot peaking at a velocity of $\sim 85 \pm 5$~km~s$^{-1}$. Its location relative to the disc ring is as expected for donor emission and thus the CaII S-wave is revealing the secondary in GW~Lib for the first time.

\begin{table*}
 \begin{minipage}{120mm}
  \caption{Velocity profile parameters on 2007 June 25 }
  \label{tab:cavelocity}
  \begin{tabular}{l c c c c}
    \hline
    Line & $\gamma$ & $K$ & $\phi_0$ & Identification \\
         &  $\mathrm{km\,s}^{-1} \pm 3 \sigma$ & $\mathrm{km\,s}^{-1}\pm 3 \sigma$ &  $\pm 3 \sigma$&\\
    \hline
    Balmer emission & $-7.6 \pm 5.6$ & $19.2 \pm 5.3$& $0.68\pm 0.04$& Accretion disc\\
    Ca~\textsc{ii} emission & $20.6 \pm 3.3$ & $19.8\pm 4.2 $& $0.73 \pm 0.04$& Accretion disc\\
    Ca~\textsc{ii} emission & $-13.1 \pm 3.7$& $82.2\pm 4.9$& $0.00 \pm 0.01$& Donor star\\
    \hline
  \end{tabular}
\end{minipage}
\end{table*}

\subsection{Emission from the accretion disc}
\label{subs:emission}

With the outburst activity decaying away, the disc contribution to the overall luminosity decreases and significant regions that are optically thin start to re-appear. These give rise to the characteristic double-peaked profile visible in both the Balmer lines and the Ca~\textsc{ii} lines. A single broad Gaussian profile fit to this feature in the phase-binned spectra together with a weighted radial velocity fit for both the Ca~\textsc{ii} and Balmer profiles at multiple nights gives an average value for the amplitude of the disc of $K_{\mathrm{acc}} = 19.2 \pm 5.3$~km~s$^{-1}$, see Table \ref{tab:cavelocity}. Similar results are obtained using multiple Gaussian line fits and diagnostic diagrams with double Gaussians. This amplitude is significantly different from the amplitude of disc related features in the April 2007 and July 2004 data. The phase shift of both Ca~\textsc{ii} and the Balmer disc coincide but are not in anti-phase with the donor star with respectively $\phi = 0.68 \pm 0.04 $ and $\phi = 0.73 \pm 0.04$. They cannot directly track the WD and thus the radial velocity amplitude is disturbed by some residual disc asymmetry and cannot be straightforwardly connected to $K_1$. 

The accretion disc in the Doppler maps of H$\beta$ appears rather featureless with little azimuthal structure, while the Ca~\textsc{ii} trail and \textsc{mem} map show lower intensity in the disc around phases $\sim 0.10$, $\sim0.35$ and $\sim 0.70$. 
The centre of emission of the discs for both maps was determined by looking at their centre of symmetry. To determine the optimal centre of symmetry of the disc, we subtracted a symmetrical component from both the Ca~\textsc{ii} and the H$\beta$ Doppler map centred at $V_x$, $V_y$ in the ranges of -100 -- +50~km~s$^{-1}$ making grid steps of 2~km~s$^{-1}$ and inspected the residuals in the lower half of the map (see also \citealt{steeghs2002}).  The best fits are found for $V_x = 0.0 \pm 5.0$~km~s$^{-1}$ and $V_y = -6.0 \pm 5.0$~km~s$^{-1}$ for the Ca~\textsc{ii} map. The fits for the H$\beta$ give identical values but with a larger error. The low $V_x$ is encouraging from the point of view of linking the disc centre with the WD at $(0,-K_1)$. The implied $K_{\mathrm{disc}}=-V_y=6$    km~s$^{-1}$ is smaller than the semi-amplitude derived using the Gaussian fits. On the other hand they are formally consistent when considering our error estimates, and both point to a small $K_1$. 

\subsection{Systemic velocity}
The binary systemic velocity ($\gamma$) is a quantity that should obviously be the same in all spectral lines, reflecting the fixed radial velocity of the binary system with respect to us. For different lines at different epochs, the systemic velocity was determined using two methods.

The first method determines the systemic velocity as derived from the radial velocity curve discussed previously, see Tables \ref{tab:velocity} and \ref{tab:cavelocity}. For the Balmer lines associated with the accretion disc $\gamma = -26 \pm 6.7$~km~s$^{-1}$ before outburst, during outburst $\gamma = 22.5 \pm 4.4$~km~s$^{-1}$ and after outburst $\gamma = -7.3 \pm 3.6$~km~s$^{-1}$. 
The narrow H$\beta$ emission during outburst has  $\gamma = -16.6 \pm 0.8$~km~s$^{-1}$. For our Ca~\textsc{ii} profiles, we found $\gamma = 20.6 \pm 3.3$ ~km~s$^{-1}$ for the disc emission and $\gamma = -13.1 \pm 3.7$~km~s$^{-1}$ for the donor star emission line. The latter is the most reliable value for $\gamma$ as the donor star has to move with the systemic velocity.

The alternative method uses the dependence of \textsc{mem}  maps on the assumed systemic velocity. Reconstructed emission spots in the map have optimal sharpness at the correct value for $\gamma$, while reconstructions at significantly different values will broaden the features and introduce possible artefacts.  Ranging $\gamma$ between -50 -- +20 km~s$^{-1}$ in steps of 5 km~s$^{-1}$ gives the best reconstruction for $\gamma = -15 \pm 5$~km~s$^{-1}$. 

Our various estimates therefore point towards a $\gamma$ of $ -15 \pm 5$~km~s$^{-1}$, in agreement with the measurements by \citet{szkody2000}.  The only outliers are the Balmer absorption line during outburst with a systemic velocity fit of $22.5 \pm 4.4$~km~s$^{-1}$ and the Ca~\textsc{ii} emission associated with the accretion disc with $\gamma = 20.6 \pm 3.3$km~s$^{-1}$. However, those fits are distorted as the resolution is not sufficient to properly resolve the donor star S-wave and the disc emission peaks near phases 0.25 and 0.75. Subtraction of the S-wave changes and fitting the disc emission with a two-Gaussian profile with an fixed offset between the Gaussians gives a lower value of $\gamma = 7 \pm 2 $~km~s$^{-1}$.

\section{System parameters}
\label{sec:systemparameters}
In the previous section we have seen the superior sensitivity of the Ca~\textsc{ii} emission lines over the commonly observed Balmer lines. Not only do they provide a much sharper view on the accretion disc emission, the key result was the presence of an emission spot from the donor star. This has provided the first ever proxy for its orbital velocity. 

Post-outburst radial velocity curves together with Doppler maps have given us estimates of the semi-amplitudes for the different components in the binary system. From both the Balmer and Ca~\textsc{ii} double-peaks, the radial velocity of the accretion disc was found to be $K_\mathrm{acc} = 19.2 \pm 5.3 $~km~s$^{-1}$. However, this value may be biased by disc asymmetries and the donor star S-wave. The observed phase-shift between the disc peaks and the donor does indeed suggest such a bias to be present, while the centre of symmetry searches returned lower $K_1$ estimates. We therefore consider $K_{\mathrm{acc}}$ from the disc peaks as an upper limit on the radial velocity of the WD: $K_1 < K_{\mathrm{acc}}$. 

The donor star is detected in the Ca~\textsc{ii} emission as a third peak on top of the accretion disc emission and gives $K_{\mathrm{em}} =  82.2 \pm 4.9$~km~s$^{-1}$. 
Considering the faintness of the low mass donor star, this is unlikely to be powered by chromospheric emission. In any case, the Ca~\textsc{ii} triplet in chromospherically active stars is generally seen in absorption (e.g. \citealt{kafka2006}) and thus cannot explain the emission seen in GW Lib. A more likely origin is photo-ionisation in the irradiated hemisphere of the donor star facing the WD. The measured radial velocity semi-amplitude is then an underestimate of the true orbital velocity of the donor star ($K_2>K_{\mathrm{em}}$) since not all of the Roche lobe contributes. We calculated the magnitude of this bias, referred to as the K-correction, by simulating the expected emission profiles from an irradiated Roche lobe filling star. Synthetic profiles were calculated for relevant binary parameters including possible shielding of the equatorial regions of the donor by a vertically extended accretion disc.

The peak intensity of the donor star S-wave does not vary significantly as a function of the orbital phase, and instead scatters around a mean value with an RMS of 8 per cent, which is consistent with photon noise.  The model profiles discussed above were also used to generate expected orbital light curves of the donor star emission as a function of binary parameters. Model light curves, assuming that the emission is optically thick and thus should be weighted with the projected area, produce orbital modulations above what is observed. However, if we consider more isotropic emission, the modulation disappears at low orbital inclinations. This optically thin model is more appropriate for emission stimulated by photo-ionisation and thus the observed lack of variability is consistent with an irradiated secondary observed at low inclinations.

We can now derive conservative limits on the radial velocities of the stellar components in the system using various methods and spectral lines. They can be summarised as $K_2 > 82.2 \pm 4.9$~km~s$^{-1}$ and $K_1 < 19.2 \pm 5.3$~km~s$^{-1}$. These limits translate into a hard upper limit for the mass ratio of $q < 0.23$ regardless of the magnitude of the K-correction or the $K_1$ overestimate. 

After the April 2007 outburst, two detections of superhumps have been reported. \citet{copperwheat2009} detect a periodicity with a period excess of $4.12$~min which implies a mass ratio of $q \simeq 0.211$ if interpreted as a superhump and the empirical superhump-excess mass ratio relation from \citet{patterson2005} is applied. This is close to our hard upper limit for the mass ratio. However, the authors themselves question the character of the humps and the implied donor stars mass in combination with the WD mass from seismology. On the other hand, \citet{kato2008} report a superhump period of $0.053925(4)$~days and a late superhump period of $0.054156(1)$~days which imply an extreme mass-ratio of $q=0.062$.

We looked at a simple model for the relation between $M_2$, $K_1$ and $K_2$ as function of $q$ for a range of $M_1$. Combining this with the above upper and lower limits for the radial velocities of respectively the WD and donor star rules out systems with $M_1 < 0.75 M_\odot$. As the system is tidally locked, the rotation period of the secondary is equal to the rotation period of the system. Emission lines are broadened by the rotation of the secondary star where the surface velocity can be approximated by looking at the co-rotating Roche Lobe surface. Comparing the measured \textsc{fwhm} of the emission line to the theoretical value as a function of the mass ratio can provide another limit on $q$ and $M_1$. The observed \textsc{fwhm} of Ca~\textsc{ii} line emission peak in each bin ranges between 1.0 and 2.2\AA~due to contamination with the double-peaked profile. The \textsc{fwhm} is very close to the resolution of the data, 0.617\AA, which equates to a $v\sin i$ of ~21~km~s$^{-1}$. The profile is thus at best marginally resolved.  From the upper limits for the mass ratio and the constraints given by the measured emission line broadening we get a rough estimate of the WD mass at $1.0 M_\odot \pm 0.25$, in agreement with the mass determination by \citet{townsley2004}. Note that their window of solutions is constrained by the UV-flux of the WD in combination with the measured distance to GW~Lib. For a parallax distance of $104 ^{+30}_{-20}$pc (from \citealt{thorstensen2003}) they calculate $M_1 = 1.03 - 1.36 M_\odot$, superseding the previous estimates by \citet{szkody2002} and \citet{thorstensen2002} that were based on a larger distance. For further analysis we will use the asteroseismologicaly suggested mass for $M_1$. 

When considering the empirical CV donor sequence of \citet{knigge2006} at the orbital period of GW~Lib, we find a typical pre/post period bounce secondary star mass of $M_2 = 0.064 \pm 0.001 M_\odot$ or $M_2 = 0.060 \pm 0.001 M_\odot$, respectively. The small difference is due to the orbital period being very close to the bounce period. Knigge also calculates the expected absolute magnitudes for GW~Lib in the infrared bands using its parallax distance, which implies a low mass secondary star of $0.080 \pm 0.005 M_\odot$.

\begin{table}
  \caption{Derived system parameters for GW Lib using $M_1$ from \citet{townsley2004}. The left hand column lists the binary parameters derived from the formal solution based on the measured limits for $K_1$ and $K_2$. The right hand column lists the parameters when combining the mass ratio inferred from superhumps with the calculated $K_2$ from the Ca~\textsc{ii} $K_{\mathrm{em}}$. }
  \label{tab:systemparameters}
  \begin{tabular}{ l l l}
\hline
     & $q \lesssim 0.23$ & $q = 0.062$ \\
\hline
$M_1\,(M_\odot)$ & $1.02$ & $1.02$ \\
$K_2\,( \mathrm{km \, s^{-1}})$ & $> 82.2$ & $100.8 \pm 7.1 $\\
$K_1\,( \mathrm{km \, s^{-1}})$ & $< 19.2$ & $6.2 \pm 0.4$\\
$M_2\,(M_\odot)$ & $< 0.23$ & $0.063$\\
$v_1\,( \mathrm{km \, s^{-1}})$ & $< 122$ & $34$ \\
$v_2\,( \mathrm{km \, s^{-1}})$ & $> 488$& $548$  \\
$i\, (^{\circ}) $ & $> 9.4$ & $10.6 \pm 0.8$\\
SpT & M3.9-M5 & M8-T  \\
\hline
  \end{tabular}
\end{table}

\subsection{Formal solution:  $q \lesssim  0.23$}
The formal scenario assumes that corrections that need to be applied to our $K_1$ and $K_2$ constraints are modest. Thus implying that the mass ratio is near its limit: $q = K_1/K_2 \lesssim 0.23$. The right hand column in Table \ref{tab:systemparameters} list the derived system parameters for this ratio.  \citet{thorstensen2002} calculated the inclination based on the \textsc{fwhm} of the shallow peaked Balmer emission lines in quiescence compared to the \textsc{fwhm} of WZ~Sge and found $i \sim 11^{\circ}$. This is in agreement with our calculations in combination with the nominal solution for $M_1 = 1.02 \, M_\odot$ and confirms that GW~Lib is indeed observed at very low inclination.  

For $q < 0.23$, the correction for $K_2$, as derived from our models, is at most $\Delta K_{\mathrm{max}} = 35 $~km~s$^{-1}$ giving a maximum value for $K_2$ of $ 118\pm 5$~km~s$^{-1}$.

The main problem with a high value for $q$ near our limit is that the secondary would then need to be a $M_2 \sim 0.23 \,M_\odot$ mass star with a  spectral type around M4-M5 (\citealt{cox2000}). In that case, spectral features should be visible in a high resolution, high signal to noise red spectrum. We cannot identify any of such in our post-outburst I-band spectra nor in the pre-outburst Magellan data. Additional issues are the observed IR magnitudes of GW~Lib which imply a faint donor and a main-sequence object near $0.23 \,M_\odot$ would not fit within the Roche-lobe. All adding to the conclusion that the mass-ratio in GW~Lib is likely not near this maximum value. 

\subsection{Superhump solution: $q=0.062$}
Measuring the late superhumps in GW Lib, \citet{kato2008} found an extreme mass ratio for the system of $q = 0.062$. Which, when combined with the WD mass from seismology and the empirical donor star mass from the \citet{knigge2006} sequence, proves an attractive solution within the boundaries of the formal scenario. 
We calculated the magnitude of the K-correction at this mass ratio using our irradiated secondary model. In order to produce the observed $K_{\mathrm{em}}=82.2\pm4.9$~km~s$^{-1}$, we find a true $K_2 = 100.8 \pm 7.1$~km~s$^{-1}$  where the error is calculated by propagating the error on $K_{\mathrm{em}}$. This error dominates over effects such as the assumed level of equatorial shielding or the optical thickness of the emission. The implied system parameters based on this assumed $q$ and calculated $K_2$ can be found in the left hand column of  Table \ref{tab:systemparameters}.

The small mass ratio scenario would correspond to a low mass donor with a M8 \citep{cox2000} or T spectral type \citep{knigge2006} if the donor is close to its main-sequence configuration. This would agree with the infrared magnitude constraints mentioned previously and fits perfectly with GW~Lib being close to the period minimum. 

Interestingly, the derived $K_1$ in this scenario is close to the measured radial velocity of the narrow Balmer emission component in outburst ($K = 3.0 \pm 1.6$~km~s$^{-1}$) suggesting that this component may trace the WD when irradiated by the accretion disc. Similar sharp emission components from the WD are seen in AM~CVn systems (\citealt{morales-rueda2003}, \citealt{roelofs2006}). However, a component arising at the surface of a WD is expected to be gravitationally red-shifted but the systemic velocity for this lines component was $-16.6 \pm 0.8$~km~s$^{-1}$ which is similar to the systemic velocity of the  system and cannot be biased by the Stark effect as is the case in AM~CVn systems.  Unfortunately the emission is only a transient feature and the constraint on its $K$ will be difficult to improve upon. The origin of the component and its possible connection with the WD thus remains unclear.

\section[]{Discussion}
\label{sec:discussion}

We presented time-resolved optical spectroscopy of GW~lib during a large number of epochs spanning before, throughout and after the 2007 outburst. We studied the long-term evolution of the spectral features tracking large changes in the accretion geometry and intensity. 

Pre-outburst spectroscopy obtained in 2004 data show clear features of the accreting WD including a detection of Mg~\textsc{ii} absorption. Here, the low inclination of GW~Lib has the advantage  over high inclination system as it gives narrow disc emission lines instead of wide profiles, potentially resolving the line from the nearby He~\textsc{i} line. Higher resolution, high signal to noise spectra around this line in quiescence could provide a dynamical trace of the accreting WD and give a directly measured mass constraint to complement the constraints indirectly obtained from seismology.

During the outburst, we initially see the optically thick accretion disc dominate through broad absorption features. Their radial velocity does not trace the WD and show semi-amplitudes of order 50 km/s. A peculiar sharp emission line component was found in the Balmer lines that is effectively stationary.
This could be some low velocity outflow in the z-direction or a component near the WD, but its mean velocity is similar to $\gamma$ whereas a layer near the high gravity WD should be gravitationally red-shifted.
As the system fades back towards quiescence, accretion powered double-peaked emission profiles appear. The absorption associated with the accretion disc continues to weaken until the system returns to a semi-quiescent state with strong double-peaked emission lines.

The Ca~\textsc{ii} triplet in GW Lib shows that these much neglected lines of CV spectra could be an interesting window to search for signs of both donor stars and accretion disc structures even in cases where the Balmer lines show no signs of the donor star whatsoever and the disc itself is barely resolved. GW~Lib is a particularly challenging object in that sense due its very low inclination and thus small projected velocities. The Ca~\textsc{ii} triplet in emission could be resolved into several components and the secondary star was discovered in emission moving in between the sharp double-peaked emission from the accretion disc. Doppler tomography and radial velocity profiles fits of the Ca~\textsc{ii} lines provide a  semi-amplitude of $K_{\mathrm{em}}=82.2 \pm 4.9$~km~s$^{-1}$ for this donor star feature and indicates a disc centre of symmetry at $K_{\mathrm{disc}} = 6 \pm 5$~km~s$^{-1}$. The disc is also visible in the H$\beta$ maps but with less detail and sharpness and no donor star contribution is seen. Contrasting these lines highlights the diagnostic advantages provided by the Ca~\textsc{ii} triplet.

Based on previous studies together with the limits on $K_1$ and $K_2$ provided in this work, the allowed range of binary parameters were explored. While our dynamical limits place a hard upper limit on the binary mass ratio of $q<0.23$, we favour a significantly lower value. A mass ratio near $q\sim0.06$ is in accordance with estimates based on the detected super-hump modulations, the constraints on the faintness, and thus mass, of the donor star and the indications that $K_1$ is very small. 
Given such a mass ratio, our measured $K_{\mathrm{em}}$ implies $K_2 = 100.8 \pm 7.1$~km~s$^{-1}$ for the donor star component when applying the relevant K-correction. The implied $K_1=6.2 \pm 0.4$~km~s$^{-1}$ is then also close to our measured disc centre of symmetry.
The combination of a WD mass near the value suggested by the pulsations and a low mass donor near the empirical sequence of an evolved CV near the period bounce appears to be consistent with the observational constraints to date.

Whether the spectral features observed several months after the tail end of the 2007 outburst are persistent remains to be established. As the WD continues to cool, donor star irradiation may be less effective in exciting the strong Ca~\textsc{ii} lines we observed. Nonetheless, high resolution spectroscopy resolving these lines at good S/N levels appears to be a viable tool to expand our knowledge of the binary parameters of short period cataclysmic variables and their faint donor stars.

\section*{Acknowledgments}

We acknowledge with thanks the variable star observations from the AAVSO International Database contributed by observers worldwide and used in this research. DS acknowledges a STFC Advanced Fellowship. We thank Perry Berlind and Mike Calkins for their assistance with obtaining the FWLO/FAST target of opportunity spectroscopy. The WHT is operated on the island of La Palma by the Isaac Newton Group in the Spanish Observatorio del Roque de los Muchachos of the Instituto de Astrofisica de Canarias. This paper includes data gathered with the 6.5 meter Magellan Telescopes located at Las Campanas Observatory, Chile operated by the Carnegie Institution of Washington and the 1.6 meter Tillinghast Telescope located at the Fred Lawrence Whipple Observatory operated by the  Smithsonian Astrophysical Observatory.

\label{lastpage}

\end{document}